\documentclass[epj]{webofc}
\usepackage[varg]{txfonts}   

\usepackage{bm}
\newcommand{\be}{\begin{equation}}
\newcommand{\ee}{\end{equation}}
\woctitle{QCD@work-International Workshop on QCD - Theory and Experiment}
\begin{document}
\title{Torsional oscillations of strange stars}
\author{Massimo Mannarelli\inst{1}\fnsep\thanks{\email{massimo@lngs.infn.it}}         
}

\institute{INFN, Laboratori Nazionali del Gran Sasso, Via G. Acitelli,
22, I-67100 Assergi (AQ), Italy}

\abstract{%
Strange stars are one of the hypothetical compact stellar objects that can be formed after a supernova explosion. The existence of these objects relies on the absolute stability of strange {\it collapsed} quark matter with respect to standard nuclear matter.  We discuss simple models of strange stars with a bare quark matter surface, thus standard nuclear matter is completely absent.
In these models  an electric  dipole layer a few hundreds Fermi thick should exist close to the star surface.  
Studying  the torsional oscillations of the electrically charged layer  we  
estimate the emitted power, finding that it is of the order of  $10^{45}$ erg/s, meaning that  these objects would be among   the brightest compact sources in the heavens. The associated relaxation times are very uncertain, with values ranging between microseconds and minutes, depending on the crust thickness. Although  part of the radiated power should be  absorbed by the electrosphere surrounding the strange star,  a sizable fraction of photons should  escape and  be detectable.}
\maketitle
\section{Introduction}
\label{intro}
\begin{figure}[h]
\centering
\includegraphics[width=10cm,clip]{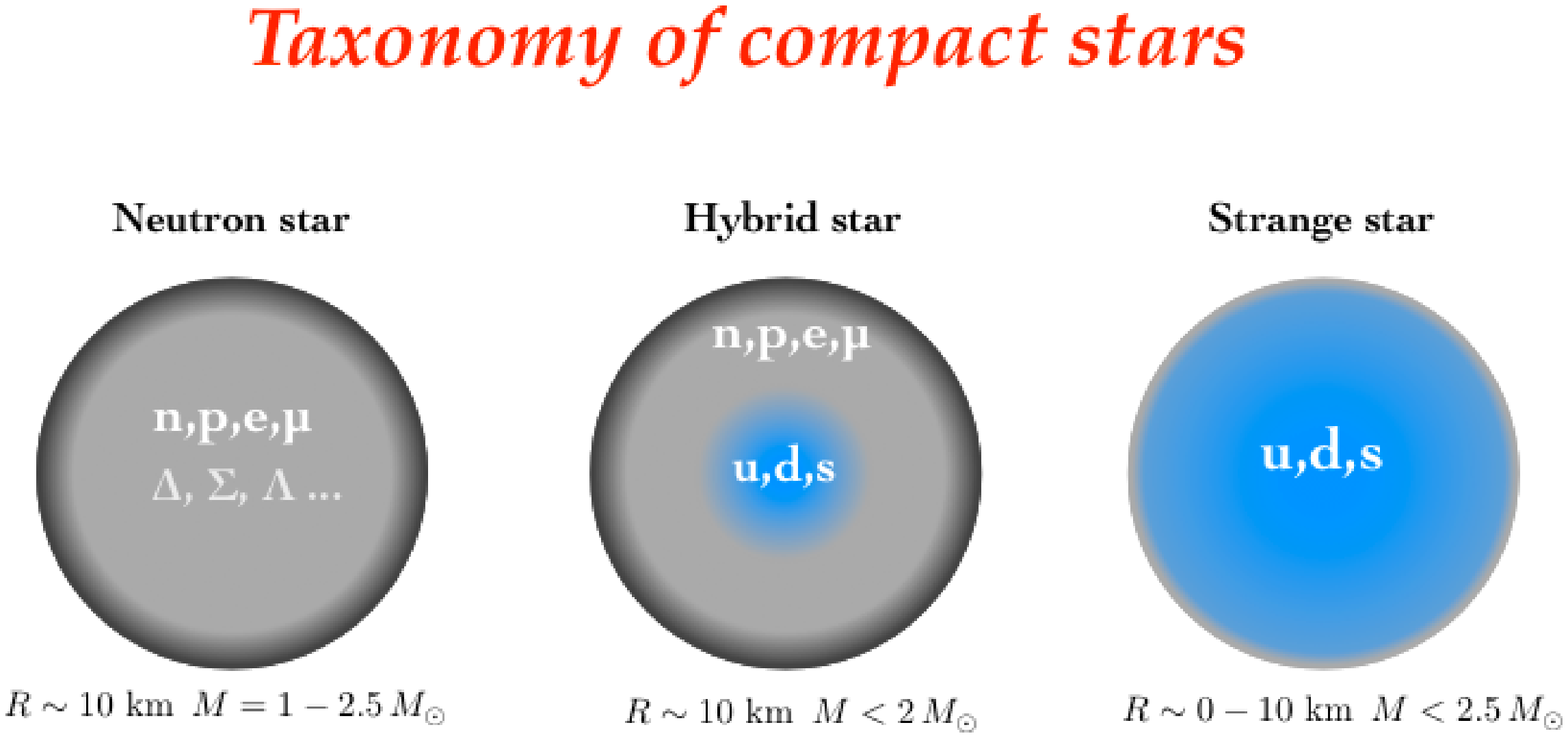}
\caption{Schematic representation of the possible realization of compact stars.}
\label{fig:compact-stars}       
\end{figure}

The properties of hadronic matter at  densities larger than the saturation nuclear  density  and at  low temperature (say $T \lesssim 1$ MeV)  are mainly unknown. 
These properties are relevant for describing the compact stellar objects (CSOs) schematically shown in Fig.~\ref{fig:compact-stars}.   The simplest, longstanding and widely studied class of CSOs are neutron stars, with all the possible hadronic variations reported in the left panel of Fig.~\ref{fig:compact-stars}. 
 The second class, central panel of Fig.~\ref{fig:compact-stars}, corresponds to the so-called hybrid stars,  having a core of deconfined quark matter enveloped in standard nuclear matter. The third class, right panel of Fig.~\ref{fig:compact-stars}, corresponds to strange stars, which are entirely made by deconfined quark matter~\cite{Alcock:1986hz, Haensel:1986qb}, see~\cite{Madsen:1998uh} for a review. The existence of CSOs having mass  between one and two solar masses, and small radii, of about $10$ km, relies on a firm observational ground. However, the identification of the observed objects with one of the CSOs reported in Fig.~\ref{fig:compact-stars} is much more uncertain.

The very existence of strange stars, relies on the assumption that standard nuclei at some density become metastable or unstable.  The favored state should correspond to a {\it collapsed state}~\cite{Bodmer:1971we} having a radial size of about $1$ fm. This apparently bizarre fact can be easily obtained by standard  nuclear physics considerations, for example see the discussion given by Blatt and Weiskoff in~\cite{blatt-weiskoff}. The reasoning goes as follows. The typical attractive  range of the  strong interaction is roughly $b \sim 1/m_\pi \sim 1.5 $ fm. Assuming that the interaction between two nucleons can be approximated  by a simple potential well of range $b$, then all nucleons should collapse in the  minimum energy state at a  distance between nucleons $\sim b/2\sim 1$fm. 

 As is well known, the collapsed state is not the observed state of nuclei;  the reason is that at very short ranges the interaction between nucleons is repulsive.  Indeed, heavy nuclei have a radius $r_\text{nucl} \simeq 1.2 \times A^{1/3}$ fm, meaning that the ground state is characterized by a matter density that is almost independent of $A$ and by a  potential energy growing as  $A$.  In other words, it is hard to  squeeze a nucleus by adding nucleons. Heavy nuclei  behave as almost incompressible liquid droplets. 
 
\begin{figure}
\centering
\sidecaption
\includegraphics[width=6cm,clip]{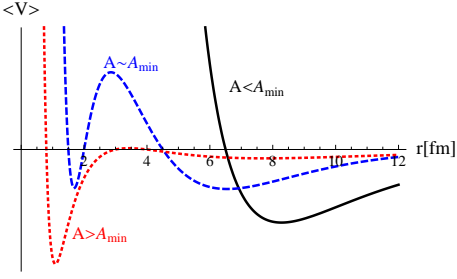}
\caption{Hypothetical behavior of the expectation value of the  two body nuclear potential with increasing baryonic number $A$. The solid line is a sketch of the standard potential energy for nuclei, with say $A \simeq 300$, having a minimum at  $r \sim 10$ fm.  The dashed line corresponds to the hypothetical behavior for   $A \sim A_\text{min}$. A local  minimum  at a distance of the order of $1$ fm corresponding to the  collapsed state has emerged. The dotted line illustrates the hypothetical behavior for $A>A_\text{min}$, with a global minimum  at $r \sim 1$ fm corresponding to the stable collapsed state.}
\label{fig:potential}       
\end{figure}

The observed saturation property of standard nuclear matter inherits from the aversion of nuclei to overlap. However, as far as I know, this property does not rely on any first principle calculation: it is an experimental fact. Thus, it might well be that a  collapsed state exists but is protected by a energy barrier and can only be accessed by  rare tunneling transitions from standard nuclear configurations.   It might happen  that with increasing baryonic number the energy barrier lowers and/or that the collapsed state becomes more energetically favored. 
In Fig.~\ref{fig:potential} we sketch a possible behavior of the average  potential energy with increasing baryonic number. The actual value of the minimum baryonic number, $A_\text{min}$, at which the conversion takes place is unknown, but it is clear that if $A_\text{min}$ exists, then for $A> A_\text{min}$  nuclei will sooner or later tunnel to the collapsed state.   If the conversion from nuclei to collapsed matter takes place at  densities close to the saturation density   and if the tunneling amplitude is sufficiently large,  then dense clumps of hadronic matter should turn to collapsed matter with no nuclear matter leftovers. In this case, the massive remnants of supernova explosions could turn into collapsed hadronic matter.

The existence of two minima in nuclear  configurations might be viewed as a case of isomerism between two configurations separated by a relatively large energy barrier. Standard examples of nuclear isomerism are fusion processes of diatomic  molecules, or fission process of heavy nuclei. It might well be that collapsed mesonic states exist  in particle physics, one example might be the $X(3872)$ resonance observed at Bell~\cite{Acosta:2003zx}. This exotic might   be a collapsed state of  four quarks~\cite{Maiani:2004vq}, see \cite{Swanson:2006st} for a review, with the  corresponding isomeric state   a $D_0 - \bar D_0^*$ molecule. In this respect,  the study of exotic hadronic states could be viewed as a study of  strongly interacting  nontrivial collapsed states.

In the collapsed state it is conceivable that the  quark content of hadrons is liberated (but this is not the only possibility, see~\cite{Bodmer:1971we}). If plenty of light quarks are present they  fill the corresponding Fermi levels  making energetically favorable the electroweak production of strange quarks. Therefore, in CSOs the  collapsed  $uds$  state could be present corresponding to  
the ground state of hadrons~\cite{Witten:1984rs}. Since in the collapsed phase matter is dense and relatively cold,
it is reasonable to expect that it is in a color superconducting phase~\cite{Rajagopal:2000wf, Alford:2007xm, Anglani:2013gfu}. The reason is that  the critical temperature of color superconductors is large, $T_{c} \simeq 1 -50$ MeV, much larger than the typical temperature of sufficiently old CSOs.

In the following we  report on the results obtained in~\cite{Mannarelli:2014ija} studying the torsional oscillations of 
strange stars having no nuclear matter crust (so-called bare strange stars~\cite{Alcock:1986hz}), with a color-flavor locked  (CFL)~\cite{Alford:1998mk} core  and a crystalline color superconducting (CCSC) crust, see \cite{Anglani:2013gfu} for a review.

\section{Equilibrium configurations}
The equation of state (EoS) of matter in the collapsed phase is unknown and cannot be determined by first principles. The reason is that even for the matter densities achievable in compact stars we expect that QCD is non perturbative.
Nonetheless, it is reasonable to expect that at sufficiently large densities  the grand potential is a function of the average baryon chemical potential, $\mu$, and a Taylor expansion  gives~\cite{Alford:2004pf}

\be\label{eq:EoS}
\Omega_{\text{QM}} = -\frac{3}{4 \pi^2} a_4 \mu^4 + \frac{3}{4 \pi^2}  a_2 \mu^2 + B_{\text{eff}}\,,
\ee
where $a_4$, $a_2$ and $B_{\text{eff}}$ are independent of $\mu$;  see~\cite{Alford:2004pf}  for a discussion of the
relevant range of values of each parameter.
In order to take into account the impact of the uncertainty of these coefficients on our results,
we consider two extreme situations, namely A ($a_4=0.7$, $a_2=(200$ MeV)$^2$ and $B_{\text{eff}}=(165$ MeV)$^4$)  and 
B ($a_4=0.7$, $a_2=0$ and $B_{\text{eff}}=(145$ MeV)$^4$).

Solving the TOV equations using the above EoS one obtains a sequence of stable configurations, see~\cite{Mannarelli:2014ija} for more details. Our reference models are - Model A, with mass of $M=1.27 M_\odot$, $R\simeq
7.1$ km and - Model B, with $M\simeq2.0 M_\odot$, $R\simeq 10.9$ km. 

\begin{figure}
\centering
\sidecaption
\includegraphics[width=5cm,clip]{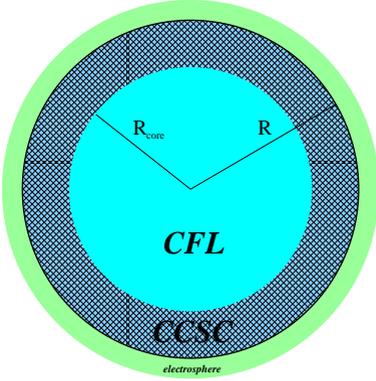}
\caption{Representation of the bare strange star model. The star is entirely made by color superconducting  matter in two different phases. The {\it core}, with radius $R_\text{core}$, consists of  color-flavor locked  color superconducting matter.
 The  {\it crust}  is made by the  crystalline color superconducting matter.  The star radius, $R$, is determined by the solution of the TOV equations with the EoS in Eq.~\eqref{eq:EoS}. We treat  the core radius, $R_\text{core}= a R$, as a free parameter. The strange star is surrounded by a cloud of electrons, the {\it electrosphere},  having a width (not in scale in the figure) of hundreds of Fermi, see Sect.~\ref{sec:charge}.}
\label{fig:star}       
\end{figure}

In both models we assume that the strange matter is actually a  color superconductor and that at a certain radial distance, $R_\text{core}=a R$ with $0 \le a \le 1$, there is a phase transition between the CFL phase  and the  three-flavor CCSC phase.  In Fig.~\ref{fig:star} we show a pictorial description of the star structure.  Since the values of $M_s$ and of the gap parameters are unknown, it is not possible to determine from first principles the radial distance at which the CFL phase turns into the CCSC phase. For this reason  we  treat $a$ as a parameter. 

One of the remarkable properties of the CCSC phase is that it is rigid. In fact what is rigid is the space modulation of the condensate. The shear modulus of the crystalline modulation was evaluated in ~\cite{Mannarelli:2007bs} and turns out to be 
\be\label{eq:nu}
\nu \simeq \nu_0\left(\frac{\Delta}{10 \text{ MeV}}\right)^2\left(\frac{\mu}{400 \text{ MeV}}\right)^2\,,
\ee
where
\be \label{eq:nu0}
\nu_0= 2.47 \frac{\text{MeV}}{\text{fm}^3}\,, 
\ee
is our reference value.  The reader is warned that the actual value of the shear modulus might differ from $\nu_0$ by a large amount because of the various approximations used in~\cite{Mannarelli:2007bs}. The  value of $\Delta$ is also uncertain,  ranging
between $5$ MeV and $25$ MeV, see~\cite{Mannarelli:2007bs,Anglani:2013gfu}.

\subsection{Charge distribution}
\label{sec:charge}
Let us now focus on the surface of the bare strange star. The surface is defined as the place where the baryonic density drops to zero. We shall now show that  a dipole charge distribution is present if the strange quark mass is sufficiently large. 

Consider first the case $M_s=0$. Then, an equal number of up, down and strange quarks is present. The reason is that  the system can minimize the pressure by transforming, by weak interactions, part of the $u$ and $d$ quarks in $s$ quarks. 
 Indeed, reducing the number of up and down quarks, reduces the corresponding Fermi energy and pressure. 
 The relevant weak processes are \be u \to d + e^+ + \nu_e \qquad u \to s + e^+ + \nu_e \qquad u + d \leftrightarrow u +s
\,. \ee  These processes will ensure that $\mu_d = \mu_u + \mu_e$ and that $\mu_d = \mu_s$.   
Since   the system has to be electrically neutral, then  $2/3 n_u -1/3 n_d -1/3 n_s - n_e =0$. For massless  strange quarks one has that  $n_u=n_d=n_s$ and then  no electrons are present.

Consider now the case  $M_s \neq 0$. As in the previous case, the weak interaction will transform part of the light quarks in strange quarks. However, less $s$ quarks are present, because they are penalized by the nonvanishing strange quark mass. Indeed, for very large $M_s$ strange quarks are absent. The charge neutralization is now achieved by the presence of electrons.  

The presence of electrons in unbalanced quark matter is the key point.
Indeed, electrons are not bound by the strong force; they are only bound by  the electromagnetic force. Therefore, they can spill outside the star surface, forming a negative charged {\it electrosphere}.  Since the surface of the star has been depleted by electrons, it becomes positively charged. The actual charge distribution can be determined by the  
 Poisson's equation 
\be\label{eq:poisson}
\frac{d^2 \phi}{d z^2} =   \sum_{i=u,d,s,e} \!\! e Q_i\, n_i(z) \,,
\ee
where $\phi$ is the electric potential,  $e Q_i$ is the charge of the species $i$ and  for simplicity we assumed  a planar  interface with $z$ measuring the distance from the quark matter discontinuity,
located at $z=0$. The star interior corresponds to $z <0$. 

The space dependence of the number densities can be  be taken into account in the local density approximation
by defining the space dependent effective chemical potential
\be\label{eq:mueffective}
\mu_i(z) = \mu_i + e Q_i \phi(z)\,.
\ee

\begin{figure}
\centering
\sidecaption
\includegraphics[width=6cm,clip]{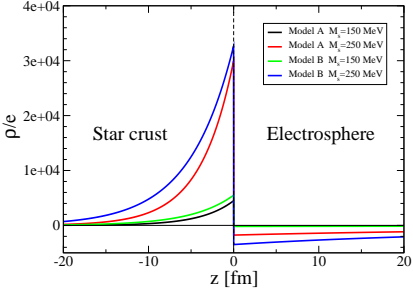}
\caption{Charge distribution close to the star surface (corresponding to $z=0$), obtained solving Eq.~\eqref{eq:poisson}. The results obtained for different models of bare strange stars and for two different values of the strange quark mass are reported.}
\label{fig:charge}       
\end{figure}

Solving the Poisson's equation  with the appropriate boundary conditions (see \cite{Mannarelli:2014ija} for more details), one can determine the potential $\phi(z)$ and feeding it in Eq.~\eqref{eq:mueffective} one determines the space dependent effective chemical potential and therefore the number densities.  The charge distribution for the various models is reported in Fig.~\ref{fig:charge}.

\section{Nonradial oscillations}
\label{sec:nonradial}
Given the equilibrium configuration studied in the previous section,
it is interesting to understand what happens when the star is put in movement. Indeed,   there is a big electric dipole moment close to the star surface  which turns out to be a powerful source of electromagnetic (EM) radiation~\cite{Mannarelli:2014ija}. In particular, we studied the EM emission associated with the torsional oscillations of the crust.

\begin{figure}
\centering
\sidecaption
\includegraphics[width=8.cm]{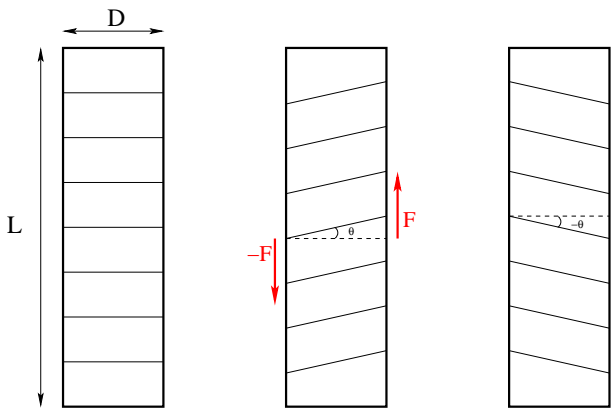}
\caption{Schematic description of the torsional oscillation of a homogeneous slab with $D \ll L$.  In the equilibrium configuration (left panel)  all the horizontal segments of the slab are parallel. A small torque applied at the surfaces of the slab slightly deforms it  tilting  the segments   (central panel). When  the applied torque vanishes (right panel), the slab starts to oscillate and the segments oscillate around the equilibrium configuration. The restoring force governing the oscillation is  the shear stress. The frequency of the oscillation is proportional to $\sqrt{\nu/(\rho D^2)}$, where $\rho$ is the matter density and $\nu$ is the shear modulus.    
 \label{fig:slabs}}
\end{figure}

The torsional  oscillations can be produced by acting with a  torque on a rigid structure, as shown in Fig.~\ref{fig:slabs} for a simple  rigid slab. When the applied forces are parallel to the sides of the slab they produce a deformation of the structure. As the external torque vanishes, the slab starts to oscillate around the equilibrium configuration.  The restoring force is proportional to the shear modulus and simple dimensional analysis gives the frequency of the small amplitude oscillations  \be\label{eq:omegaslab} \omega \propto \frac{1}D\sqrt{\frac{\nu}{\rho}}\,,\ee where $D$ is the thickness of the slab. The slab can be thought as a local approximation of the CCSC crust, and therefore we expect that the frequency of the crust torsional oscillations has the same  qualitative dependence on $\nu$, $\rho$ and the crust thickness, $D=R-R_\text{core}$ as in Eq.~\eqref{eq:omegaslab}. The detailed analysis of \cite{Mannarelli:2014ija} and sketched below shows that this is indeed the case.

Our interest in  the torsional oscillations is clearly due to the fact that in the CCSC phase the shear modulus is extremely large and can therefore sustain   oscillations of large amplitude and high frequency. 

Torsional oscillations are characterized by a displacement vector field $\bm \xi$ such that
\be
\bm \nabla \cdot \bm \xi =0 \qquad \text{and} \qquad  \xi_r =0\,.
\ee
The first condition implies that there is no volume compression. The second condition implies that there is no radial displacement. In spherical coordinates the eigenmodes of the displacement vector can be written as \be\label{eq:displacement}
\xi_{nl}^r =0 \qquad \xi_{nl}^\theta = 0 \qquad \xi_{nl}^{\phi} = \frac{W_{nl}(r)}{r \sin\theta} \frac{\partial P_l(\cos \theta)}{\partial \theta} e^{i \omega_{nl} t}\,,
\ee
where $l$ is the angular momentum, $n$ is the principal quantum number,  $P_l(\cos \theta)$ are the Legendre polynomials and we assumed that the eigenmode has frequency $\omega_{nl}$. The study of these  oscillations in the Newtonian approximation gives fairly good results, see \cite{1988ApJ...325..725M};  anyway, general relativity corrections are certainly smaller than the large uncertainties of the various parameters of our models.  

For definiteness, we  assume that the only excited mode is the one  with $l=1$ and $n=1$, corresponding to the lowest energy nontrivial mode of a nonrotating star.  The corresponding  oscillation frequency turns out to be
\be\label{eq:w1}
\omega_{11} \simeq 0.06  \left( \frac{\nu}{\nu_0} \right)^{1/2}\!\!\!\left( \frac{\delta R}{1 \text{km}} \right)^{-1} \left(\frac{\rho_R}{\rho_0} \right)^{-1/2} \text{MHz}\,,
\ee 
where $\delta R= (1-a)R$ and $\rho_0=10^{15} \text{g/cm}^{3}$. Note that 
overtones with $n > 1$ and/or $l > 1$, having higher frequencies, could  as well be excited by the external agency triggering the oscillation of the crust. 

\begin{figure}
\centering
\sidecaption
\includegraphics[width=7.cm]{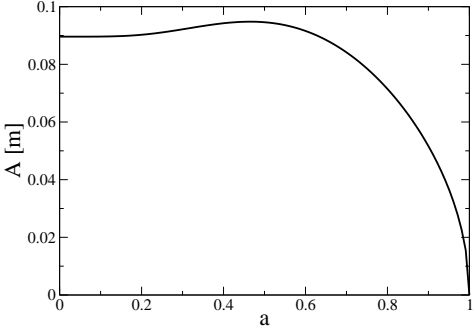}
\caption{Horizontal displacement of the torsional oscillation evaluated at the star surface as a function of $a=R_\text{core}/R$, see Eq.~\eqref{eq:W11}.
 \label{fig:amplitude}}
\end{figure}

The amplitude of the crust oscillations is determined by the amount of energy that the triggering agency can store in the mode. Various possible  triggers exist. For definiteness we considered nonradial oscillations  triggered by a stellar glitch. In this case it is natural to assume that   a fraction $\alpha <1$ of the energy of a glitch excites the torsional oscillations.  Assuming that only the $l=1, n=1$ mode is excited we have that
\be\label{eq:glitch}
\alpha E_{\text{glitch}} = \frac{\rho_R \omega_{11}^2}{2}   \int  |\bm \xi_{11}|^2 dV \,,
\ee
where  we consider as a reference value $E_{\text{glitch}}^{\text{Vela}} = 3\times10^{-12} M_\odot$ as estimated for the giant Vela glitches.  Of particular relevance for us is the amplitude of the oscillation at the star surface, because it  determines the oscillation of the positive  electric charge. The amplitude of the displacement, defined in Eq.~\eqref{eq:displacement}, can be expressed as
\be\label{eq:W11}
W_{11}(R) = A(a)  \left( \frac{\nu}{\nu_0}\right)^{-1/2} \left(\frac{R}{10 \text{km}}\right)^{-1/2}\left(\frac{\alpha E_{\text{glitch}}}{E_{\text{glitch}}^{\text{Vela}} }\right)^{1/2}\,,\,
\ee
where $A(a)$ is the function reported in Fig.~\ref{fig:amplitude}. 
The amplitude of the oscillation is extremely large, leading to a very high electric current at the surface of the star. We estimated the emitted power by a magnetic dipole oscillator in~\cite{Mannarelli:2014ija}. For $a > 1/2$  an approximated solution is 
\be\label{eq:Pa}
P(a) \simeq 6.4 \times 10^{40} (1-a)^{-5}   \left(\frac{\nu}{\nu_0} \right)^2  \left(\frac{\rho_R}{\rho_0} \right)^{-3} \left(\frac{R}{10 \text{km}}\right)^{-1} \left(\frac{\alpha E_{\text{glitch}}}{E_{\text{glitch}}^{\text{Vela}} }\right) \left( \frac{Q_+}{Q}\right)^2 \text{erg/s}\,,
\ee
where we have averaged over time and considered as a reference value for the surface charge density $Q= 10^5$ MeV$^3$fm, see  \cite{Mannarelli:2014ija} for more details. One remarkable property is that the radiated power increases with increasing $a$:  the thinner the crust, the larger is the radiated power.

\section{Conclusions}
\label{sec:conclusions}
Strange stars are one of the possible realizations of compact stellar objects. Since the pioneering work by {\it Alcock et al.} \cite{Alcock:1986hz}, it is known that these stars have en electric dipole layer with a very high electric field. 
Assuming that the star has a crust of CCSC matter  
we estimated the energy  emitted by an oscillation of this dipole layer triggered by a glitch. The  emitted power is extremely large, being of the order of $10^{41}$ erg/s. Although the electrosphere will screen a large fraction of the released photons, reducing the emissivity by about one order of magnitude~\cite{Mannarelli:2014ija}, these stellar objects  are certainly among the brightest compact sources in the sky. 

Various astrophysical sources are known to emit electromagnetic radiation at high energy. 
Two notable examples are the Rotating Radio Transients (RRTs) and the giant magnetar x-rays flares.

The RRTs have observed frequencies of the order of GHz, a duration of few milliseconds and an extremely large energy flux, see~\cite{McLaughlin:2005eq, Lorimer:2007qn, Thornton:2013iua}. Their sources are presently unknown. Although the results of our calculations seem to be in qualitative agreement with the above properties, for proposing a candidate model we have to refine our treatment of the EM emission. As an example,    we assumed the coherent emission of EM radiation,  certainly overestimating   the emitted power at GHz frequencies.

Giant x-rays flares of magnetars~\cite{Strohmayer:2005ks}  are challenging phenomena, which can be hardly explained  by strange stars with no crust~\cite{Watts:2006hk}. The standard explanation of these flares is indeed related with the seismic vibrations of the nuclear crust triggered by a starquake. Typical frequencies are of the order of hundreds of Hz  at most and the emitted luminosities is extremely large, ranging between  $10^{44} - 10^{46}$ erg/s. The measured decay time is of order of minutes. In our model,  oscillations of hundreds of  Hz can be reached only if the shear modulus is sufficiently small, making it comparable with the one of  standard  nuclear crusts,   and if the CCSC crust is  sufficiently thick, say of the order of a few kilometers. Needless to say that such a strange star would pretty well  mimic a standard neutron star.   

A lot of work on strange stars has been done so far. Although we believe that this class of compact stars cannot be ruled out by present astronomical observations, it is certainly true that no strong evidences of their existence have been accumulating in these years. In our analysis we pointed out a new possible observable, associated with the EM emission due to the torsional oscillations of the crystalline color superconducting crust.

\end{document}